\documentclass[aps,pra,reprint,twocolumn,superscriptaddress,showkeys,showpacs]{revtex4}
\usepackage{amsfonts}
\usepackage{amssymb}
\usepackage{bm, bbm, dsfont}
\usepackage[dvips]{graphicx}
\usepackage[T1]{fontenc}
\usepackage[latin2]{inputenc}
\usepackage{amsmath}
\usepackage{latexsym}

\usepackage{color}

\def\dt{{\rm d}\,}

\newcommand{\ket}[1]{| #1 \rangle}
\newcommand{\bra}[1]{\langle #1 |}

\def\duzomniejsze{<\kern-.7mm<}
\def\duzowieksze{>\kern-.7mm>}

\def\textbf#1{{\bf #1}}
\def\be{\begin{equation}}
\def\ee{\end{equation}}
\def\ben{\begin{eqnarray}}
\def\een{\end{eqnarray}}
 \def\beqa{\begin{eqnarray}}
\def\eeqa{\end{eqnarray}}
\def\eea{\end{array}}
\def\bea{\begin{array}}
\newcommand{\bei}{\begin{itemize}}
\newcommand{\eei}{\end{itemize}}
\newcommand{\bee}{\begin{enumerate}}
\newcommand{\eee}{\end{enumerate}}

\def\hcal{{\cal H}}

\def\1{\openone}

\def\tr{{\rm Tr}}

\def\>{\rangle}
\def\<{\langle}

\def\dt#1{{{\kern -.0mm\rm d}}#1\,}
\def\squareforqed{\hbox{\rlap{$\sqcap$}$\sqcup$}}
\def\qed{\ifmmode\squareforqed\else{\unskip\nobreak\hfil
\penalty50\hskip1em\null\nobreak\hfil\squareforqed
\parfillskip=0pt\finalhyphendemerits=0\endgraf}\fi}

\newtheorem{lemma}{Lemma}
\newtheorem{theorem}[lemma]{Theorem}
\newtheorem{main result}[lemma]{Main result}
\newtheorem{proposition}[lemma]{Proposition}
\newtheorem{definition}{Definition}

\newtheorem{fact}[lemma]{Fact}

\def\bep{\begin{proposition}}
\def\eep{\end{proposition}}
\def\bel{\begin{lemma}}
\def\eel{\end{lemma}}

\def\bet{\begin{theorem}}
\def\eet{\end{theorem}}
\def\bed{\begin{definition}}
\def\eed{\end{definition}}
\def\bef{\begin{fact}}
\def\eef{\end{fact}}

\begin{document}

\title{Quantum Origins Of Objectivity}

\author{R.~Horodecki}
 \affiliation{Institute of Theoretical Physics and Astrophysics, University of Gda\'nsk, 80-952 Gda\'nsk, Poland}
 \affiliation{National Quantum Information Centre in Gdan\'sk, 81-824 Sopot, Poland}
\author{J.~K.~Korbicz}
  \affiliation{Institute of Theoretical Physics and Astrophysics, University of Gda\'nsk, 80-952 Gda\'nsk, Poland}
  \affiliation{National Quantum Information Centre in Gdan\'sk, 81-824 Sopot, Poland}
\author{ P.~Horodecki}
\affiliation{Faculty of Applied Physics and Mathematics, Gda\'nsk University of Technology, 80-233 Gda\'nsk, Poland}
\affiliation{National Quantum Information Centre in Gdan\'sk, 81-824 Sopot, Poland}

\date{\today}

\begin{abstract}
In spite of all of its successes, quantum mechanics leaves us with a central problem: How does Nature create a ''foot-bridge'' from fragile quanta to  
the objective world of everyday experience?
Here we find that a basic structure  within quantum mechanics that leads to the 
perceived objectivity is a,  so 
called, spectrum broadcast structure. We uncover this basing  on 
minimal assumptions, without referring to any dynamical details or a concrete model. 
More specifically, working formally within the decoherence theory setting with multiple environments (known as quantum Darwinism),
we show how  a crucial for quantum mechanics notion of non-disturbance due to Bohr 
and  a natural definition of objectivity lead to  a canonical  structure of a quantum system-environment state, 
reflecting objective information records about the system stored in the environment.
\end{abstract} 
\maketitle

\section{Introduction}

The emergence of objective world from quanta has been a long 
standing problem, already present from the very dawn of quantum mechanics \cite{old_bohr,old_heis,modern}. 
One of the most promising approaches is the decoherence theory, based on a system-environment paradigm \cite{Zeh,Schloss}: 
a quantum system is considered interacting with its environment. It recovers, under certain conditions, a classical-like 
behavior of the system alone in some preferred frame, singled out by the interaction and called a pointer basis \cite{ZurekPRD}, and explains it through information 
leakage from the system into the environment. However, decoherence theory is silent on how 
comes that in the classical realm information is redundant \cite{ZurekNature}: same record can exist in a large number of copies and can be independently 
accessed by many observers and many times. Or more basically: Since quanta cannot be cloned \cite{no_cloning} 
and information redundancy is---from the observers' measurements perspective---at the heart of objectivity, then what quantum process lies at the foundations of the objective classical world ?

Recently, a crucial step was made in
a series of works (see e.g. \cite{ZurekNature, ZwolakZurek}) introducing  quantum
Darwinism idea.  Its essence is that in more realistic 
environments, composed of many independent fractions,  decoherence leads to an appearance of multiple copies of system's state in the environment, accessible to independent observers.
Although presenting a convincing physical picture, there is no general, model-independent justification of such claims apart from studies under the strict conditions 
of specific models, e.g. spin-$1/2$ systems \cite{Zurek_spins}
or an illuminated sphere \cite{sfera_zurek}. However, even those studies still do not
present totally convincing arguments within the models themselves, as they are based on a scalar information-theoretical condition 
and, so called, "partial information plots", which are known to be only
a necessary condition for objectivity but its sufficiency is still unknown.

Here we take a more fundamental and rigorous position,
based solely on what as for now provides the most basic description of Nature: A quantum state (see also \cite{generic}).
More precisely, we \emph{derive} from first principles a universal,  objectivity-carrying structure of quantum states, using a general
approach, independent of any dynamics (much like e.g. the S-matrix theory in the quantum field theory \cite{Cao}):  
Looking at the post-interaction system-environments state, we ask what properties should it have to reflect the objectivity.  Surprisingly, 
the answer comes with a help of Bohr's notion of non-disturbance \cite{Bohr, Wiseman}, which originally used to defend 
the quantum \cite{EPR,Bohr}, here, ironically, defines the classical.
It is obtained  through, what we call, a \emph{spectrum broadcast structure}, which precisely pin-points the 
distributed character of information and makes  it essentially classical. 
 We finally illustrate our approach on one of the emblematic examples of the decoherence theory: 
A dielectric sphere illuminated by photons \cite{sfera,sfera_zurek,myPRL}. 
It must be mentioned that in the quantum Darwinism literature there appeared similar 
quantum state structures (so called "branching states"). However, they have been at best 
tacitly \emph{postulated} \cite{ZurekNature, ZwolakZurek, ZurekPRA06}, if at all explicitly mentioned.
Our results allow to understand an intimate connection between the perceived objectivity, a specific structure of quantum states,
and information broadcasting.

\section{General Theorem}\label{gen}

\subsection{Basic Definitions And The Main Result}\label{results}

We first define the central concepts of our study and state the main result.
The basis of our work is the following definition of an objective state \cite{ZurekNature, Ollivier}:
\begin{definition}[Objectivity]\label{obj}
A state of the system $S$ 
exists objectively if many observers can find out the state of $S$ independently, and without perturbing it.
\end{definition}
As it stands, the above definition is rather informal and has to be made more rigorous.
For example the key concept of perturbation has to be made precise. We do it in the next Subsection.

Second key concept of our study is a \emph{spectrum broadcast structure}, defined below.
\begin{definition}[Spectrum Broadcast Structure]\label{sp}
Spectrum broadcast structure is the following form of a joint state of the central system $S$ and a collection of sub-environments $E_1,\dots,E_{fN}$ 
(denoted by $fE$):
\begin{equation}\label{br2}
\varrho_{S:fE}=\sum_i p_i \ket i_S\bra i\otimes \varrho^{E_1}_i\otimes\cdots\otimes \varrho^{E_{fN}}_i,
\end{equation}
where $\{\ket i\}$ is some basis in the system's space, $p_i$ are probabilities, and 
all states $\varrho^{E_k}_i$ are perfectly distinguishable:
\be\label{distng}
\varrho^{E_k}_i\varrho^{E_k}_{i'}=0\ \text{ for all}\ i\ne i' \text{ and for all } k=1,\dots,fN. 
\ee
\end{definition}
All the nomenclature will be clarified in the next Subsection.
This is a special  form  of, so called, classical-classical states, which
have been  introduced as a counterpart of separable states in the 
context of quantification of quantum correlations \cite{Oppenheim, CC}.
It has first appeared in a context of quantum channels in \cite{my}.

The main result of this work is establishing of an intimate connection between these two concepts.
The two pivotal assumptions we will use are that of \emph{Bohr non-disturbance} \cite{Bohr,Wiseman} and \emph{strong independence}.
The first has been formulated in \cite{Wiseman} (see Section \ref{clar}), while by the strong independence we mean that
\emph{the only correlation between the environments should be the common information about the system}.
In other words, conditioned by the information about the system, there should  be  no correlations between the environments. 
This is in a sense an idealization, which we use since we are interested in the information flow only between the system and 
each of the environments, but not between the environments themselves.
Under these central assumptions (together with some auxiliary ones), we prove in Section \ref{proof} the following Theorem:
\begin{theorem}\label{thm2} Assume that a system undergoes a full decoherence. Then
the appearance of a spectrum broadcast structure is a necessary and sufficient condition
for  objectivity in the sense of Definition \ref{obj}:
\begin{eqnarray}
\begin{array}{c}
\text{Objective Existence} \bm{\Leftarrow}\text{Spectrum Broadcast Structure}\\
\left(\begin{array}{c} \text{Objective}\\ 
\text{Existence}\end{array}\right)+\left(\begin{array}{c} \text{Strong}\\ 
\text{Independence}\end{array}\right)\bm{\Rightarrow} \left(\begin{array}{c} \text{Spectrum}\nonumber\\ 
\text{Broadcast}\nonumber\\
\text{Structure} \end{array}\right).
\end{array}
\label{impl}
\end{eqnarray}                    
\end{theorem}

\subsection{Formalization of Definition \ref{obj}}\label{clar}

Here we put Definition \ref{obj} into a physical frame and make it as precise as possible, which is 
the hardest work. As the most suitable, we formally choose decoherence theory
with multiple environments \cite{ZurekNature}: The quantum system of interest 
$S$ interacts with multiple environments $E_1, \dots ,E_N$ (denoted collectively as $E$), also modeled as quantum systems. 
The environments (or their collections) are assumed to be \emph{macroscopic} and are monitored by independent observers \cite{ZwolakZurek}. 
The motivation behind such a choice is that in real-life situations there is always present some interaction with the environment 
(unless very special conditions are met) and we, the observers, have usually access only to a small portion of it, each to a different.
However, as we will see in what follows, no assumptions on the dynamics will be needed.
In fact, we may forget about the dynamics altogether and pose a more general question:
which multipartite system-environment states reflect objectively existing state of the system in the sense of Def.~\ref{obj}?

We only assume that the system-environment interaction is such that it leads to a full decoherence.
The standard, or even paradigmatic, case corresponding to the latter is a physical situation when there exists a time scale $\tau_D$, called the
decoherence time, such that asymptotically for interaction times $t\gg \tau_D$: i) there emerges in the system's Hilbert space
a unique, stable in time pointer basis $\{\ket i\}$; ii) the reduced state of the system 
$\varrho_S$ becomes stable and diagonal in the pointer basis:
\be \label{decoh}
\varrho_S\equiv\tr_E\varrho_{S:E}\approx\sum_i p_i \ket{i}\bra i, 
\ee
where $p_i$'s are some probabilities and 
by $\approx$ we will always denote asymptotic equality in the deep decoherence limit $t/\tau_D\to\infty$.
However, it should be stressed, that while we usually will mean the latter situation, our derivation of 
the structure of objectivity,
covers also possible situations when the process happens in finite time. 
We assume here the above explained \emph{full decoherence}, so that the system 
decoheres in a basis rather than in higher-dimensional pointer superselection sectors.
This is, because we want to consider the full objectivisation of a given quantum
 degree of freedom,  rather than a partial one.
Clearly, the environment must be of a large dimension  to 
have a big informational capacity, needed to carry highly redundant records about the decohered  system $S$. 
Moreover, some loss of information is needed (and of course in fact happens in the reality), as otherwise
there will be no decoherence, and we assume that some of the environments pass unobserved. 
The observed environments $E_1, \dots ,E_{fN}$ we denote by  $fE$ (depending on the context). 


Next, we specify the observers. Apart from the environmental ones, 
we also allow for a (possibly hypothetical) direct observer, who can measure the system $S$ directly. Such an observer is needed as a reference, 
to verify that the findings of the environmental observers are the same as if one had a direct access to the system. 

The word ''find out'' we translate as the observers performing  von Neumann---as perfectly repeatable contrary to 
the generalized---measurements on their subsystems. 
It should be stressed here that von Neumann measurement, with its repeatability property, has been chosen since 
we identify  the spectrum broadcast structure \emph{as the paradigmatic, ideal structure of the state, responsible for objectivity}.
 Indeed, this is the  the object to which any real physical state should be compared with, if we want to know, whether 
the objectivity in more or less approximate sense (in terms of a state trace distance) takes place.
(Note that it can be compared with the ideal singlet as the target state of quantum distillation or 
the ideal channel, in the case of coding theory, to which the outputs of real protocols or physical 
situations are compared).

By the ''independence'' requirement of Def. \ref{obj}, there can be no correlations between them. 
Consequently, the global von Neumann measurement, resulting from the individual local observers' measurements, 
must be fully product:
\ben\label{prod}
\Pi^{M_S}_i\otimes\Pi^{M_1}_{j_1}\otimes\cdots\otimes\Pi^{M_{fN}}_{j_{fN}},
\een
where $M_S,M_1,\dots,M_{fN}$ denote measurements on $S, E_1,\dots,E_{fN}$ respectively 
and all $\Pi$'s are mutually orthogonal Hermitian projectors, $\Pi^{M_k}_j\Pi^{M_k}_{j'\ne j}=0$.
The observers so determine the probabilities $p_i$ of $\ket i$ in (\ref{decoh}) 
(they must know the pointer basis $\{\ket i\}$, as if not, they 
would not know what information they get is all about). 
As explained before Theorem \ref{thm2}, we will actually demand more by assuming
the strong independence.

The most crucial clarification needed in Definition \ref{obj} is to make precise the word "perturbing".
We apply here Bohr's notion of 
non-disturbance \cite{Bohr, Wiseman},
according to which given local measurements on the subsytems are non-disturbing
if they leave the \emph{whole} joint state invariant (after forgetting the results).
This is a realistic mathematical idealization of a repetitive information 
extraction---a crucial prerequisite for objectivity. 

We recall that Bohr's non-distrurbance was formulated in order to save
the completeness of quantum theory against the famous  Einstein-Podolsky-Rosen argument \cite{EPR}.
Bohr argued \cite{Bohr} that the EPR notion of mechanical non-disturbance (which amounts to the no-signaling
principle \cite{Wiseman}) was too restricted and a broader notion was needed.
Hence, accepting the completeness of quantum theory, as we do for the purpose of this work, one
is forced to accept Bohr's notion of non-disturbance.

Finally, the independent measurements will typically reveal inconsistent information about the system
(see however \cite{generic}). Indeed, allowing for general correlations may lead to a disagreement: 
if one of the observers measures first, the ones measuring afterwards may find outcomes depending on the result of the first measurement. 
Thus, we add to Def. \ref{obj} an obvious \emph{agreement requirement}: ''...many observers can find out 
\emph{the same} state of $S$ independently...''.

\subsection{Proof of Theorem \ref{thm2}}\label{proof}

We are now ready for a proof of Theorem \ref{thm2} with the additional assumption
of \emph{strong independence}, explained in Section \ref{results}. We first prove that spectrum broadcast structure from Definition \ref{sp}
is a sufficient condition for an objectively existing state of the system, in the sense of Definition \ref{obj}. 
Indeed, from (\ref{br2}) projections on $\ket i$ and on the disjoint supports of  $\varrho^{E_k}_i$ 
constitute the non-demolition measurements. Performing them independently, 
all the observers will repeatedly detect the same index $i$ with probabilities $\{p_i\}$, without Bohr-disturbing the joint $S:fE$ state,
thus making the sate $\ket i$ exist objectively in a sense of Def. \ref{obj} (cf. \cite{ZurekPRA}).

We now prove in the opposite direction. We assume 
the decoherence has taken place (cf. \cite{generic}).
Crucial here is the Bohr's non-disturbance condition from Subsection \ref{clar}.
Together with 
the product structure (\ref{prod}), it
implies that on each subsystem $S,E_1, . . .E_{fN}$ there exists a non-demolition von Neumann measurement,
leaving the whole asymptotic state $\varrho_{S:fE}(\infty)$ of the system and the observed environment invariant 
(the symbol $\infty$ stands here  either for the  $t/\tau_D\to\infty$ asymptotic, or as mentioned before, for 
any time scale, may be finite, after which the objectivity structure emerges). 
For $S$ it is defined by the projectors on $\ket i$. For the environments we 
allow for higher-rank projectors $\Pi^{M_k}_j$,$k = 1, . . . ,fN$, not necessarily spanning the whole space, 
as the environments can have inner degrees of freedom not correlating to $S$.

Consequently, the total joint probability of the results of the Bohr non-disturbing measurements is given by :
\be\label{p_joint}
p_{ij_1\dots j_{fN}}\equiv\tr\left[\ket i\bra i\otimes \Pi^{M_1}_{j_1}\otimes\cdots\otimes\Pi^{M_{fN}}_{j_{fN}}\varrho_{S:fE}(\infty)\right].                                                                                                                                             
\ee 
Now the agreement requirement from Subsection \ref{clar}
leads to a natural conclusion:
\be\label{agree}
p_{ij_1\dots j_{fN}}\ne 0\  \text{  iff  } \ i=j_1=...=j_{fN}.
\ee
Let us more formally show it,
considering for simplicity only two observers.
If one of them measures first and gets a result $i$, then the joint conditional state becomes
$\varrho_{|i}=(1/p_i)(\Pi_i\otimes{\bf 1})\varrho (\Pi_i\otimes{\bf 1})$, $p_i\equiv\tr(\Pi_i\otimes{\bf 1}\varrho)$
and the subsequent measurement by the second observer will yield results $j$ with conditional probabilities
$p_{j|i}=(1/p_i)\tr(\Pi_i\otimes\Pi_j\varrho)$.
If for some $i$, $p_{j|i}p_{j'|i}\ne 0$ for $j\ne j'$, 
then comparing their results after a series of measurements at some later moment, the observers
will be confused as to what exactly the state the system $S$ was: with the probability $p_{j|i}p_{j'|i}$ 
the second observer will obtain different states $j\ne j'$, while the first observer measured the same state $i$.
One would not have the observers' findings objective, unless for every $i$ there exists only one $j(i)$ such that $p_{j(i)|i}\ne 0$ (actually
$p_{j(i)|i}=1$, which follows from the normalization $\sum_i p_{i|j}=1$, so that the distributions $p_{\cdot|i}$ are all deterministic).
Reversing the measurement order and applying the same reasoning, we obtain that  
for every $j$ there can exist only one $i(j)$ such that $\tilde p_{i(j)|j}\ne 0$, where by the Bayes theorem
$\tilde p_{i|j}= p_{j|i}p_i/\tilde p_j$, $\tilde p_j\equiv\tr({\bf 1}\otimes\Pi_j\varrho)$.
These two conditions imply that the joint probability $p_{ij}=p_i\delta_{ij}$ (after an eventual renumbering).
Applying the above argument to all the pairs of indices, one obtains (\ref{agree}).
This means that the environmental Bohr-nondisturbing measurements must be perfectly correlated with the pointer basis. 

Hence, after forgetting the results, the asymptotic, post-measurement state $\varrho^M_{S :fE}(\infty)$ reads:
\ben
&& \varrho^M_{S:fE}(\infty)\equiv\sum_{i,j_1,\dots , j_{fN}}p_{ij_1\dots j_{fN}}\varrho_{ij_1\dots j_{fN}}^{S:fE}(\infty)\nonumber\\
&&=\sum_i \ket i\bra i\otimes{\bf \Pi}_i\,\varrho_{S:fE}(\infty)\,\ket i\bra i\otimes {\bf \Pi}_i,
\een
where ${\bf \Pi}_i\equiv\Pi^{M_1}_{i}\otimes\cdots\otimes\Pi^{M_{fN}}_{i}$.

Now we are ready for the \emph{key step}: we impose the relevant form of the Bohr-nondisturbance condition:
\ben\label{maxcorr}
\sum_i \ket i\bra i\otimes{\bf \Pi}_i\,\varrho_{S:fE}(\infty)\,\ket i\bra i\otimes {\bf \Pi}_i=\varrho_{S:fE}(\infty),
\een
whose only solution \cite{Wiseman} are Classical-Quantum (CQ) states \cite{QC}:
\ben\label{wism}
\varrho_{S:fE}(\infty)=\sum_i p_i \ket i\bra i\otimes{\bf R}^{fE}_i,
\een
where $p_i$ are identified with the probabilities from Eq.~(\ref{decoh}) and ${\bf R}_i^{fE}$ are some residual states
in the space of the observed environments with mutually orthogonal supports: ${\bf R}_i^{fE}{\bf R}_{i'\ne i}^{fE}=0$.
Hence, ${\bf R}^{fE}_i$ are perfectly distinguishable through the assumed non-disturbing
measurements ${\bf \Pi}_i$, projecting on their supports.

Finally, let us look at the residual states ${\bf R}_i^{fE}$ in (\ref{wism}). The demand of the independent 
ability to determine the state of $S$, already used in (\ref{prod}), completed with the strong independence condition
(cf. Section \ref{gen}) leads to the following: Once one of the observers finds a particular result $i$, the resulting conditional state should be fully product.
Since the direct observer is already uncorrelated by (\ref{wism}), this implies that:
\be\label{Ri}
{\bf R}_i^{fE}=\varrho^{E_1}_i\otimes\cdots\otimes \varrho^{E_{fN}}_i.
\ee
and  $\varrho^{E_k}_i$ must be perfectly distinguishable for each  $E_k$, cf.  (\ref{distng}),
since by (\ref{maxcorr}) for any $k$ it holds: $\Pi^{M_k}_i\varrho^{E_k}_i\Pi^{M_k}_i=\varrho^{E_k}_i$  and 
$\Pi^{M_k}_i\Pi^{M_k}_{i'\ne i}=0$. This finishes the proof.

Some remarks are in order. First, in the course of the proof we have formulated
a broader class of independent environments, in a way paradigmatic in quantum information 
theory \cite{0way}: the environments are \emph{independent} if and only if the environmental
observers may produce  the states (\ref{Ri},\ref{distng}), exploting
only local operations (equivalent to local trace preserving maps), i.e.
independent environments are those ones that simulate strong independence from 
the perspective of a specific resource (the class of  local operations). 

Second, the meaning of the Theorem \ref{thm2} is that it provides 
\emph{an ideal reference structure for objectivity}---the broadcast structure (\ref{br2}).
Any other non-ideal situation should be compared to that  broadcast structure
no matter what figure of merit is taken. On the level of the states, this 
must be the trace norm which has the clear probabilistic  interpretation, where the
degree of objectivity is just the trace norm distance to the broadcast state.
The transition:
\be
\text{\emph{initial $S:E$ state}} \longrightarrow \text{\emph{spectrum broadcast structure (\ref{br2})}} \nonumber
\ee
identifies a basic process, called here \emph{state information broadcasting}, responsible 
for an appearance of the perceived objectivity. Formally, it involves 
broadcasting of a part of information about the system---the spectrum of its state after the decoherence, 
$\text{Sp}\varrho_S\equiv\{p_i\}$, into the environments and is thus similar to 
quantum state  \cite{broadcasting} and spectrum \cite{my} broadcasting.
Condition (\ref{distng}) forces the correlations in (\ref{br2}) to be entirely classical
and thus the detailed structures of $\varrho^{E_k}_i$ become irrelevant for the correlations.
From (\ref{br2}) and (\ref{distng}) it follows that under a suitable convergence:
\be\label{qd}
I\left[\varrho_{S:fE}(\infty)\right]=H_S \ \text{ for every fraction }f,
\ee
where $I(\varrho_{AB})\equiv S_{\text{vN}}(\varrho_A)+S_{\text{vN}}(\varrho_B)-S_{\text{vN}}(\varrho_{AB})$ is the quantum mutual information, 
$S_{\text{vN}}(\varrho)\equiv -\tr(\varrho\log \varrho)$ stands for the
von Neumann entropy, and $H_S\equiv S_{\text{vN}}[\varrho_S(\infty)]=H(\{p_i\})$ is the entropy of the decohered  state (\ref{decoh}).
Condition (\ref{qd}), postulated as a sufficient condition for objectivity in quantum Darwinism model,
has a clear meaning in the classical information theory \cite{ThomasCover}: every fraction $f$ carries the same information $H_S$ about the 
system---the latter is redundantly encoded in the environment. However, in the quantum world its sense remains unclear (see the next Section). Here, (\ref{qd}) follows automatically from
the deeper structure (\ref{br2}).

\section{Discussion Of The Entropic Objectivity Condition As A "Witness" For Objectivity}\label{disc}
Here we show a potential problem with the entropic objectivity condition (\ref{qd}).
as a sufficient condition for objectivity (see e.g. Refs.~\cite{ZurekNature,ZwolakZurek} and references therein).
Although our example below \emph {is not} fully conclusive, we argue that at this moment neither is
the reasoning of quantum Darwinism studies.

Condition (\ref{qd}) has been shown to hold in several models, including the illuminated sphere \cite{sfera_zurek,myPRL}
and spin baths \cite{Zurek_spins}.
For finite times $t$, the equality (\ref{qd}) is not strict and holds within some error $\delta(t)$, which defines the 
\emph{redundancy} $R_\delta(t)$ as the inverse of the smallest fraction of the environment $f_{\delta(t)}$, 
for which $I[\varrho_{S:f_{\delta(t)} E}(t)]=[1-\delta(t)]H_S$. 
When satisfied, (\ref{qd}) implies that the mutual information between the system
and the environment fraction is a constant function of the fraction size $f$ (up to an error $\delta$ for finite times)
and the plot of $I$ against $f$ exhibits a characteristic plateau, called the \emph{classical plateau} (see e.g. Ref.~\cite{ZurekNature}). 
The appearance of this plateau has been heuristically explained in the quantum Darwinism literature 
as a consequence of the \emph{redundancy}: classical information about the system exists in many copies in the environment fractions and 
can be accessed independently and without perturbing the system by many observers, thus leading to objective existence of a state of $S$ \cite{ZurekNature}.
That would be for sure so in the classical information setting: 
the condition (\ref{qd}) is there equivalent to a perfect correlation of both systems \cite{ThomasCover}, i.e.
for every $f$ the environment fraction has a full information about the system and indeed this information thus exists objectively
in the sense of our definition.

But in the quantum world the situation may be different  and the condition (\ref{qd}) 
alone may not be sufficient to guarantee objectivity, 
due to a wholistic nature of quantum correlations \cite{Michal}.
It is clear that the spectrum broadcast states (\ref{br2}) 
satisfy (\ref{qd}), but there may also be entangled states satisfying it, 
thus violating the spectrum broadcast form, derived as a necessary condition for objectivity. 
As a simple example in favour of such a statement consider the following state of two qubits,
where one is the system $S$ and the second the environment $E$:
\be
\varrho_{S:E}\equiv p P_{\left(a\ket{00} + b\ket{11}\right)} + (1-p)P_{\left(a\ket{01}+ b\ket{10}\right)}, 
\ee
where $P_\psi\equiv\ket\psi\bra\psi$, $p \neq 1/2$, $a=\sqrt{p}$ and $b=\sqrt{1-p}$. Then the partial state
$\varrho_S=\tilde p \ket{0}\bra{0}+(1-\tilde p)\ket{1}\bra{1}$, $\tilde p\equiv pa^2+(1-p)b^2$ is diagonal 
in the basis $\ket{0},\ket{1}$ and moreover $S_{\text{vN}}(\varrho_S)=S_{\text{vN}}(\varrho_{S:E})\equiv h(\tilde p)$ 
(the binary Shannon entropy \cite{ThomasCover}), so that a form of the entropic  
condition holds: $I(\varrho_{S:E})=S_{\text{vN}}(\varrho_S)=H_S$, 
$H_S=h(\tilde p)$, but the systems are nevertheless entangled, which one verifies directly through
the PPT criterion \cite{PPT}.

The above example is of course not conclusive, as there is only one environment,
but it suggest that the functional condition (\ref{qd}) in principle might indeed be not sufficient to show objectivity, as
defined in the main text. We leave this, in general uneasy, question open for further sesearch.
In the above context it can be already seen what is the  paradigmatic shift with respect to the earlier works on 
 decoherence  and quantum Darwinism models we propose here: it is the pivotal observation govering our approach that 
the core object of the analysis should be \emph{a derived structure of the full quantum state} of the system $S$ and the observed environment $fE$,  rather than the partial state of the system only (Decoherence Theory) or information-theoretical functions (quantum Darwinism).

\section{Spectrum Broadcast Structure In The Illuminated Sphere Model}
We exemplify the general findings  from Section \ref{gen} on one of the central models 
of decoherence (see e.g. \cite{sfera,sfera_zurek, myPRL}): a dielectric sphere 
illuminated by photons (for the details see Appendix \ref{sphere}). 
We show that in the course of the evolution  a broadcast state (\ref{br2}) is asymptotically formed 
in this model, assuming for simplicity pure environments (see \cite{myPRL} for a more general analysis).
The sphere is initially in a state without a well defined position
(e.g. in $\ket{\psi^S_0}=(\ket{\vec x_1}+\ket{\vec x_2})/\sqrt 2$).
Photons scatter elastically and slightly differently depending on where the sphere is, 
but this difference is vanishingly small for each individual scattering: If the observed fraction is too small, 
the post scattering states $\ket{\Psi^{mic}_i}\equiv {\bf S}_i\ket{\vec k_0}$ (${\bf S}_i$ are the scattering matrices) 
become identical in the thermodynamic limit:
$\langle\Psi^{mic}_2|\Psi^{mic}_1\rangle\equiv \langle\vec k_0|{\bf S}_2^\dagger {\bf S}_1\vec k_0\rangle\xrightarrow{\text{therm.}}1$
and the joint post scattering state approaches effectively a product: $(\sum_{i=1,2}p_i\ket{\vec x_i}\bra{\vec x_i})
\otimes\ket{\Psi^{mic}}\bra{\Psi^{mic}}^{\otimes \mu}$, where probabilities $p_i\equiv|\langle\psi_0^S|\vec x_i\rangle|^2$
form the spectrum of the decohered state (cf. (\ref{decoh})). The  photons thus force the sphere to be in a definite position $\vec x_i$  
with the probability $p_i$, but the observed fraction carries no information about it (a \emph{product phase}; see Appendix \ref{phases}).
\begin{figure}[t]
\begin{center}
\includegraphics[scale=0.34]{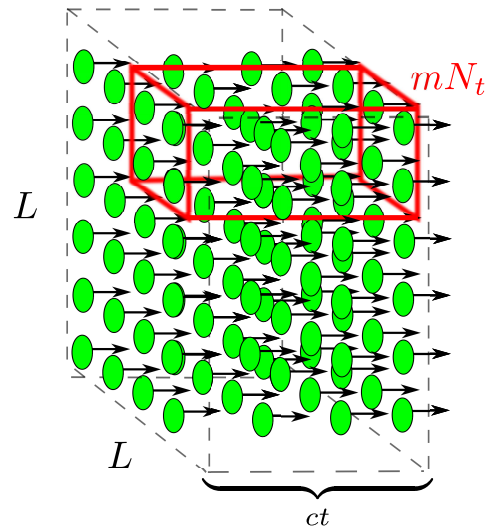}
\caption{ 
(Color online) Coarse-graining of the photonic environment. The photons (green) scattered in time $t$ are grouped into
$M$ equal macroscopic fractions $mN_t$. Only one fraction (bounded by the red cage) is shown; $L$
is the edge of an artificial box used for quantization (and removed later; see Appendix \ref{sphere}).
The macro-fractions may be thought of as representing a sensitivity of the photon detectors (e.g. an eye) and
their exact  size, $mN_t$, is irrelevant--scaling with the total photon number $N_t$ suffices.
\label{division}}
\end{center}
\end{figure} 

However, when grouped into macroscopic fractions, the photons become almost perfectly resolving. 
Imagine we divide all the photons scattered up to time $t$, $N_t$, into $M$ macro-fractions of $mN_t$, $0<m<1$,
photons, Fig.~\ref{division}. Then the \emph{macroscopic} post scattering states 
$\ket{\Psi^{mac}_i(t)}\equiv ({\bf S}_i\ket{\vec k_0})^{\otimes mN_t}$,
become asymptotically perfectly distinguishable:
\be\label{ort_main}
\left|\langle\Psi_2^{mac}(t)|\Psi_1^{mac}(t)\rangle\right|\xrightarrow{\text{therm.}}\text e^{-\frac{m}{\tau_D}t},
\ee
where $\tau_D$ is the decoherence time \cite{sfera,sfera_zurek}. If we observe $fM$, $0<f<1$,
macro-fractions out of $M$, then the joint post scattering state has asymptotically the spectrum 
broadcast structure (\ref{br2}): 
\ben
&&\varrho_{S:fE}(0)=\varrho^S_0\otimes\underbrace{\varrho_0^{mac}\otimes\dots\otimes\varrho^{mac}_0}_{fM} 
\xrightarrow[\text{therm.}]{t\gg\tau_D}\varrho_{S:fE}(\infty)=\nonumber\\
&&\sum_{i=1,2}p_i\ket{\vec x_i}\bra{\vec x_i}
\otimes\underbrace{\ket{i^{mac}}\bra{i^{mac}}\otimes\dots\otimes\ket{i^{mac}}\bra{i^{mac}}}_{fM},\label{sib}
\een
where $\ket{i^{mac}}\equiv\ket{\Psi_i^{mac}(\infty)}$ emerges, due to (\ref{ort_main}), as the non-disturbing environmental basis
in the space of each macro-fraction. Eq.~(\ref{sib}) identifies the state information broadcasting
process: the information about the sphere's localization, $\{p_i\}$, 
is redundantly transfered into the environment and becomes available in multiple copies through the measurements
in $\{\ket{i^{mac}}\}$. The process consists of: i) decoherence \cite{sfera} and ii) orthogonalization (\ref{ort_main}), and defines a \emph{broadcasting phase}, see Appendix \ref{phases}, 
corresponding to the classical plateau of  \cite{sfera_zurek}.
From Fannes-Audenaert \cite{FAd} and Alicki-Fannes \cite{FannesAlicki} inequalities, the entropic condition (\ref{qd})
follows as a consequence of (\ref{sib}) (see Appendix \ref{condition}). Finally, if all the photons are observed, the post-scattering state 
maintains the full \emph{quantum} correlation with the system
and $I[\varrho_{S:fE}(\infty)]=I_{max}$ (a \emph{full information phase}).

\section{Discussion}
In conclusion, based on an universal approach, independent of any 
dynamics or a concrete model, we have identified the primitive state information broadcasting process 
responsible for an emergence of the perceived objectivity (for a possible loosening of some of our assumptions see \cite{generic}).
Our main result (Thm. \ref{thm2}) suggests that the states of the form (\ref{sib}) are notoriously formed
in Nature. In a laboratory, this can be in principle directly verified via e.g. quantum state tomography \cite{tomo}.
Moreover, it naturally leads to a view that in fact there may be no "quantum-to-classical transition"---what we perceive
as "classical", e.g. objective information, may be merely a reflection of some specific properties of the underlying quantum states,
like the spectrum broadcast structure; a view further strengthened by \cite{ont}. 

There appears to be a deep connection between the non-signaling principle and objective existence
in the sense of Definition \ref{obj}: the core fact that it is at all possible for 
observers to determine \emph{independently} the classical state
of the system is guaranteed by the non-signaling principle: $\tr({\bf 1}_S\otimes \Pi_E\varrho_{S:E})=\tr_E(\Pi_E\varrho_E)$.
There is no contradiction with the Bohr-nondisturbance, as the latter is a strictly \emph{stronger} condition
than the non-signaling \cite{Wiseman} (this is the core of Bohr's reply \cite{Bohr} to EPR). 
In fact, the above  connection reaches deeper than quantum mechanics. In a general theory, where it is
possible to speak of probabilities $p(ij|MN)$ of obtaining results $i,j$ when performing measurements $M,N$ (however defined),  
whatever the definition of objective existence may be, the requirement of the
\emph{independent} ability to locally determine probabilities by each party seem indispensable. 
This is guaranteed in the non-signaling theories, where all $p(ij|MN)$'s have well defined marginals.
In this sense non-signaling seems a \emph{prerequisite of cognition}. 
In this context, we also believe that our approach to objectivity will open a new perspective on the celebrated Bell Theorem
\cite{Bell}. These connections will be the subject of a further research.

The emergence of redundantly encoded information in the structure of quantum states may also shed new light on the 
life phenomenon. Since self-replication of the DNA information is indispensable for the existence of life,    
it cannot be excluded that the state information broadcasting may indeed open a ''classical window'' 
for life processes within quantum mechanics \cite{Wigner}.

{\bf Acknowledgements} 
We thank W. H. Zurek and C. J. Riedel for discussions and comments and M. Piani for discussions on strong independence. 
P.H. and R.H. acknowledge discussions with K. Horodecki, M. Horodecki, and K. \.Zyczkowski. 
This research is supported by ERC Advanced Grant QOLAPS and  
National Science Centre project Maestro DEC-2011/02/A/ST2/00305.

\appendix

\section{Technical Details The Illuminated Sphere Model For  Pure  Environments}\label{sphere}
\subsection{Description of the model}
Here we present a detailed derivation of the spectrum broadcast structure (\ref{br2})
in the illuminated sphere model for pure environments (cf. \cite{myPRL} for a more general situation). 
We first recall the basics of the model, following the usual treatment 
(see e.g. Refs.~\cite{sfera,GallisFleming,sfera_zurek,RiedelZurek}).
The system $S$ is a sphere of radius $a$ and relative permittivity $\epsilon$, 
bombarded by a constant flux of photons, which constitute the multiple environments and decohere the sphere.
The sphere can be located only at two positions: $\vec x_1$ or $\vec x_2$, so that
effectively its state-space is that of a qubit $\hcal_S\equiv \mathbb C^2$ with a preferred orthonormal (due to the mutual exclusiveness) 
basis $\ket{\vec x_1}$, $\ket{\vec x_2}$, which will become the pointer basis.
This greatly simplifies the analysis, yet allows the essence of the effect to be observed. 
The sphere is sufficiently massive, compared to the energy of the radiation, 
so that the recoil due to the scattering can be totally neglected and 
photons' energy is conserved, i.e. the scattering is \emph{elastic}.
 
The environmental photons are assumed not energetic enough to individually resolve the sphere's displacement $\Delta x\equiv|\vec x_2-\vec x_1|$:
\be\label{soft}
k\Delta x\ll 1,
\ee
where $\hbar k$ is the characteristic photon momentum.
Otherwise, each individual photon would be able to resolve the position of the sphere 
and studying multiple environments would not bring anything new.
On the technical side, following the traditional approach \cite{sfera,GallisFleming,sfera_zurek,RiedelZurek}, we describe the photons 
in a simplified way using box normalization: 
we assume that the sphere and the photons are enclosed in a large box of edge $L$
and volume $V=L^3$ and photon momentum eigenstates $\ket{\vec k}$ obey periodic boundary conditions. 
Although a more rigorous treatment was developed in Ref.~\cite{HornbergerSipe} with well localized photon states, 
we choose this traditional heuristic approach as,
at the expense of a mathematical rigor, it allows to expose the physical situation more clearly, without unnecessary mathematical details 
(we remark that the findings of  Ref.~\cite{HornbergerSipe} agree with the previous works using box normalization \cite{Adler}).
After dealing with formally divergent terms, we remove the box through the thermodynamic limit 
(signified by $\cong$) \cite{sfera_zurek,RiedelZurek}: 
\be\label{thermod}
V\to\infty, N\to\infty, \frac{N}{V}=\text{const}, 
\ee
that is we expand the box  and add more photons, keeping the photon density constant, as 
the relevant physical quantity is the radiative power, proportional
to $N/V$. The thermodynamic limit is crucial in the sense that it 
defines micro- and macroscopic regimes, which will turn to be qualitatively very distinct.

The detailed dynamics of each \emph{individual} scattering is irrelevant---the individual scatterings
are treated asymptotically in time. The interaction time $t$ enters the model differently, thought the number of scattered photons.
It may be called a ''macroscopic time''.  
Assuming photons come from the area of $L^2$ at a constant rate $N$ photons per volume $V$ per unit time, 
the amount of scattered photons from $t=0$ to $t$ is:
\be\label{Nt}
N_t\equiv L^2\frac{N}{V}ct,
\ee
where $c$ is the speed of light. Throughout the calculations we work with a fixed time $t$ and
pass to the asymptotic limit $t/\tau_D\to \infty$ (signified by $\approx$ or $\infty$) at the very end.

Since multiphoton scatterings can be neglected and all the photons are treated equally 
(\emph{symmetric environments}), the effective sphere-photons interaction up to time $t$ 
is of a controlled-unitary form:
\be\label{U}
U_{S:E}(t)\equiv\sum_{i=1,2}\ket{\vec x_i}\bra{\vec x_i}\otimes \underbrace{{\bf S}_i\otimes\dots\otimes {\bf S}_i}_{N_t},
\ee
where (assuming  translational invariance of the photon scattering) 
${\bf S}_i\equiv {\bf S}_{\vec x_i}=e^{-i\vec x_i\cdot \vec{\hat k}} {\bf S}_0 e^{i\vec x_i\cdot \vec{\hat k}}$ 
is the scattering matrix 
when the sphere is at $\vec x_i$, ${\bf S}_0$ is the scattering matrix when the sphere is
at the origin, and $\hbar\vec{\hat k}$ is the photon momentum operator.
Due to the elastic scattering, ${\bf S}_i$'s have non-zero matrix elements only between the states $\ket{\vec k}$ 
of the same energy $\hbar c |\vec k|$.
In the sector (\ref{soft}) the interaction (\ref{U}) is vanishingly small at the level of each {\it individual} photon \cite{RiedelZurek}:  
in the thermodynamic limit ${\bf S}_1\cong{\bf S}_2$  (in a suitable sense we clarify later), 
and hence $\sum_i\ket{\vec x_i}\bra{\vec x_i}\otimes {\bf S}_i\cong {\bf 1}\otimes {\bf S}$. 
Surprisingly, this will not be true for macroscopic groups of photons.
We also note that unlike in the previous treatments \cite{sfera, GallisFleming, sfera_zurek, RiedelZurek, HornbergerSipe}, 
already at this moment we explicitly include in the description \emph{all} the photons scattered up to the fixed time $t$.
Finally, the preferred role of the basis $\ket{\vec x_i}$ is already singled out now by the form of the interaction (\ref{U}) \cite{ZurekNature}. 

The initial, pre-scattering ''in'' state, is as usually assumed a \emph{full product}: 
\be\label{init}
\varrho_{S:E}(0)\equiv\varrho^S_0\otimes(\varrho^{ph}_0)^{\otimes N_t},
\ee
with $\varrho^S_0$ having coherences in the preferred basis $\ket{\vec x_i}$ and $\varrho^{ph}_0$
some initial states of the photons (the environments are by assumption symmetric).
Next, we introduce a crucial \emph{environment coarse-graining} \cite{ZurekNature}:
the full environment (i.e. all the $N_t$ photons) is divided into a number of \emph{macroscopic fractions}, 
each containing $mN_t$ photons, $0\le m \le 1$. By \emph{macroscopic} 
we will always understand ''scaling with the total number of photons $N_t$''. 
By definition, these are the environment fractions accessible to the independent observers. 
Such a division may seem artificial and arbitrary, as e.g. the choice of $m$ is unspecified. 
However, observe that in typical situations detectors used to monitor fractions of the environment, 
e.g. eyes, have some minimum detection thresholds---some minimum amount of radiative energy
delivered in a given time interval is needed to trigger the detection. Each macroscopic fraction $mN_t$ 
is meant to reflect that detection threshold. Its concrete value (the fraction size $m$) 
is for our analysis irrelevant---it is enough that it scales with $N_t$. 
This coarse-graining procedure is analogous to the one used e.g. in the description of liquids:
each point of a liquid (a macro-fraction $m$ here) is in reality composed of a suitable large number of 
microparticles (individual photons). It is also employed in mathematical approach to von Neumann
measurements using, so called, macroscopic observables (see e.g. Ref.~\cite{Sewell} and the references therein).

Thus, we divide the detailed initial state of the environment $(\varrho^{ph}_0)^{\otimes N_t}$
into $M\equiv 1/m$ macroscopic fractions:
\ben
\underbrace{\varrho^{ph}_0\otimes\dots\otimes\varrho^{ph}_0}_{N_t}&=&\underbrace{\varrho^{ph}_0\otimes\dots\otimes\varrho^{ph}_0}_{mN_t}\otimes\dots\otimes
\underbrace{\varrho^{ph}_0\otimes\dots\otimes\varrho^{ph}_0}_{mN_t}\nonumber\\
&\equiv& \underbrace{\varrho_0^{mac}\otimes\dots\otimes\varrho^{mac}_0}_{M},\label{init_mac}
\een
where $\varrho_0^{mac}\equiv (\varrho^{ph}_0)^{\otimes mN_t}$ is the initial state of each macroscopic fraction 
(\emph{macro-state} for brevity).

\subsection{Dynamical formation of broadcast structure}
After all the $N_t$ photons have scattered, the asymptotic (in the sense of the scattering theory) 
''out''-state  $\varrho_{S:E}(t)\equiv U_{S:E}(t)\varrho_{S:E}(0)U_{S:E}(t)^\dagger$, is given from Eqs.~(\ref{U},\ref{init},\ref{init_mac}) by
\ben
&&\varrho_{S:E}(t)=\nonumber\\
&&\sum_{i=1,2}\langle\vec x_i |\varrho^S_0\, \vec x_i\rangle\ket{\vec x_i}\bra{\vec x_i}\otimes\underbrace{\varrho_i^{mac}(t)\otimes\dots\otimes\varrho_i^{mac}(t)}_M\label{SE1}\\
&&+\sum_{i\ne j}\langle\vec x_i |\varrho^S_0\, \vec x_j\rangle\ket{\vec x_i}\bra{\vec x_j}\otimes\underbrace{\left({\bf S}_i\varrho^{ph}_0 {\bf S}_j^\dagger\right)^{\otimes mN_t}\otimes\dots}_M\label{SE2}
\een
where 
\be\label{rho_i}
\varrho_i^{mac}(t)\equiv\left({\bf S}_i\varrho^{ph}_0 {\bf S}_i^\dagger\right)^{\otimes mN_t}, \, i=1,2. 
\ee

In order for the decoherence to take place, some of the environment must be traced out.
In the current model it is important that the forgotten fraction must be \emph{macroscopic}:
we assume that $fM$, $0\leq f\leq 1$ out of all $M$ macro-fractions of Eq.~(\ref{init_mac}) are observed, 
while the rest, $(1-f)M$, is traced out. 
The resulting partial state reads 
(cf. Eqs.~(\ref{SE1},\ref{SE2})):
\ben\label{SfE}
&&\varrho_{S:fE}(t)=\sum_{i=1,2}\langle\vec x_i |\varrho^S_0\, \vec x_i\rangle\ket{\vec x_i}\bra{\vec x_i}\otimes\left[\varrho_i^{mac}(t)\right]^{\otimes fM}\label{i=j}\\
&&+\sum_{i\ne j}\langle\vec x_i |\varrho^S_0\, \vec x_j\rangle\left(\tr {\bf S}_i\varrho^{ph}_0 {\bf S}_j^\dagger\right)^{(1-f)N_t}\ket{\vec x_i}\bra{\vec x_j}\otimes\nonumber\\
&&\qquad\qquad\qquad\qquad\qquad\qquad\otimes\left({\bf S}_i\varrho^{ph}_0 {\bf S}_j^\dagger\right)^{\otimes fN_t}.
\label{i ne j}
\een

We finally demonstrate that in the soft scattering sector (\ref{soft}), 
the above state is asymptotically of the broadcast form (\ref{br2}) 
by showing that in the deep decoherence regime $t\gg \tau_D$ two effects take place:
\begin{enumerate}
\item The coherent part $\varrho_{S:fE}^{i\ne j}(t)$ given by Eq.~(\ref{i ne j}) vanishes in the trace norm:
\be\label{znikaogon}
||\varrho_{S:fE}^{i\ne j}(t)||_{\text{tr}}\equiv\tr\sqrt{\left[\varrho_{S:fE}^{i\ne j}(t)\right]^\dagger\varrho_{S:fE}^{i\ne j}(t)}\approx 0.
\ee
\item The post-scattering macroscopic states $\varrho_i^{mac}(t)$ (cf. Eq.~(\ref{rho_i})) become perfectly distinguishable:
\be\label{nonoverlap}
\varrho_1^{mac}(t)\varrho_2^{mac}(t)\approx 0,
\ee
or equivalently using the generalized overlap \cite{Fuchs}:
\ben
&&B\left[\varrho^{mac}_1(t),\varrho^{mac}_2(t)\right]\equiv\nonumber\\
&&\quad\quad\quad\equiv\tr\sqrt{\sqrt{\varrho^{mac}_1(t)}
\varrho^{mac}_2(t)\sqrt{\varrho^{mac}_1(t)}}\approx 0,\label{nonoverlap_norm}
\een
despite of the individual (microsopic) states becoming equal in the thermodynamic limit.
\end{enumerate}
The first mechanism above is the usual decoherence of $S$ by $fE$---the suppression of coherences in the preferred basis $\ket{\vec x_i}$. 
Some form of quantum correlations may still survive it, 
since the resulting state (\ref{i=j}) is generally of a Classical-Quantum (CQ) form \cite{QC}. 
Those relict forms of quantum correlations are damped by the second mechanism: the asymptotic   
perfect distinguishability (\ref{nonoverlap}) of the post-scattering macro-states $\varrho_i^{mac}(t)$. 
Thus, the state $\varrho_{S:fE}(\infty)$ becomes of the spectrum broadcast form (\ref{br2})
for the distribution:
\ben\label{pi}
p_i=\langle\vec x_i |\varrho^S_0\, \vec x_i\rangle.
\een

We demonstrate the mechanisms (\ref{znikaogon},\ref{nonoverlap}), and hence
a formation of the broadcast state (\ref{br2}), for pure initial environments: 
\ben\label{pure}
\varrho_{ph}^0\equiv\ket{\vec k_0}\bra{\vec k_0},\ k_0\Delta x\ll 1, 
\een
i.e. all the photons come from the same direction
and have the same momenta $\hbar k_0$, $k_0\equiv|\vec k_0|$, satisfying (\ref{soft}). 
To show (\ref{znikaogon}), observe that $\varrho_{S:fE}^{i\ne j}(t)$, defined by Eq.~(\ref{i ne j}), is of a simple form in the basis $\ket{\vec x_i}$:
\be\label{matrix}
\varrho_{S:fE}^{i\ne j}(t)=\left[\begin{array}{cc} 0 & \gamma C\\
                                                                         \gamma^*C^\dagger & 0\end{array}\right],
\ee
where $\gamma\equiv\langle\vec x_1 |\varrho^S_0\, \vec x_2\rangle(\tr {\bf S}_1\varrho^{ph}_0 {\bf S}_2^\dagger)^{(1-f)N_t}$ and
$C\equiv ({\bf S}_1\varrho^{ph}_0 {\bf S}_2^\dagger)^{\otimes fN_t}$. Since ${\bf S}_i$'s are unitary and $\varrho^{ph}_0\geq 0$, $\tr\varrho^{ph}_0=1$,
we obtain:
\ben
&&||\varrho_{S:fE}^{i\ne j}(t)||_{\text{tr}}=\nonumber\\
&&|\gamma|\tr\left({\bf S}_1\varrho^{ph}_0 {\bf S}_1^\dagger\right)^{\otimes fN_t}+|\gamma|\tr\left({\bf S}_2\varrho^{ph}_0 {\bf S}_2^\dagger\right)^{\otimes fN_t}\\
&&=2|\langle\vec x_1 |\varrho^S_0\, \vec x_2\rangle|\left|\tr {\bf S}_1\varrho^{ph}_0 {\bf S}_2^\dagger\right|^{(1-f)N_t}\label{|i ne j|}
\een
The decoherence factor $|\tr {\bf S}_1\varrho^{ph}_0 {\bf S}_2^\dagger|^{(1-f)N_t}$ for the pure case (\ref{pure}) has been 
extensively studied before (see. e.g. Refs.~\cite{sfera, GallisFleming, sfera_zurek, RiedelZurek, HornbergerSipe}). Let us briefly recall the main 
results. Under the condition (\ref{soft}) and
using the classical cross section of a dielectric sphere in the dipole approximation $k_0a\ll1$, 
one obtains in the box normalization:
\ben\label{psipsi}
&&\langle\vec k_0|{\bf S}_2^\dagger {\bf S}_1\vec k_0\rangle=1+i\frac{8\pi\Delta x k_0^5 \tilde a^6}{3L^2} \cos\Theta\nonumber\\
&&-\frac{2\pi\Delta x^2 k_0^6 \tilde a^6}{15L^2}
\left(3+11\cos^2\Theta\right)+O\left[\frac{(k_0\Delta x)^3}{L^2}\right],
\een
where $\Theta$ is the angle between the incoming direction $\vec k_0$ and the displacement vector $\vec{\Delta x}\equiv\vec x_2-\vec x_1$
and $\tilde a\equiv a [(\epsilon-1)/(\epsilon+2)]^{1/3}$. This implies:
\ben
&&\left|\tr {\bf S}_1\varrho^{ph}_0 {\bf S}_2^\dagger\right|^{(1-f)N_t}=\left|\langle\vec k_0|{\bf S}_2^\dagger {\bf S}_1\vec k_0\rangle\right|^{(1-f)N_t}\cong\nonumber\\
&&\left[1-\frac{2\pi\Delta x^2 k_0^6 \tilde a^6}{15L^2}
\left(3+11\cos^2\Theta\right)\right]^{L^2(1-f)\frac{N}{V}ct}\label{linia2}\\
&&\xrightarrow{\text{therm.}}\text e^{-\frac{(1-f)}{\tau_D}t}.\label{decay_ogon}
\een
In the second line above we used Eq.~(\ref{psipsi}) up to the leading order in $1/L$; in the last line we removed the box normalization
through the thermodynamical limit (\ref{thermod})  and thus obtained the decoherence time \cite{sfera_zurek,RiedelZurek}:
\be
{\tau_D}^{-1}\equiv\frac{2\pi}{15}\frac{N}{V}\Delta x^2 c k_0^6 \tilde a^6 \left(3+11\cos^2\Theta\right).
\ee
Eqs.~(\ref{|i ne j|},\ref{decay_ogon}) imply that $||\varrho_{S:fE}^{i\ne j}(t)||_{\text{tr}}\leq 2\text e^{-(1-f)t/\tau_D}|\langle\vec x_1 |\varrho^S_0\, \vec x_2\rangle$, 
since the sequence $(1+x/N)^N$ is monotonically increasing.
As a result, whenever we forget a macroscopic fraction of the environment ($f<1$),
the resulting coherent part $\varrho_{S:fE}^{i\ne j}(t)$ decays in the trace norm exponentially, with the characteristic
time $\tau_D/(1-f)$. This completes the first step (\ref{znikaogon}).
\begin{figure}[t]
\begin{center}
\includegraphics[scale=0.32]{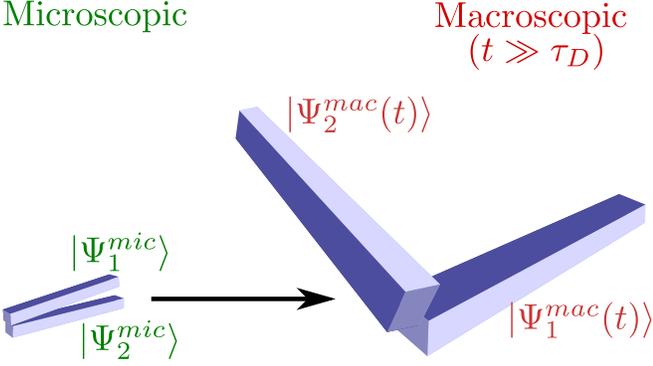}
\caption{(Color online) \emph{Orthogonalization of macroscopic states}. 
At the \emph{microscopic} level, the individual post-scattering states $\ket{\Psi^{mic}_i}\equiv {\bf S}_i\ket{\vec k_0}$, 
corresponding to the sphere being at $\vec{x}_i$ 
(represented by the small solid slabs on the left)
become identical in the thermodynamic limit (cf. Eq.~(\ref{micro})) and hence completely indistinguishable.
They carry vanishingly small amount of information
about the sphere's localization, which is due to the assumed weak coupling between the sphere and each individual
environmental photon (\ref{soft}). On the other hand, the collective 
states of \emph{macroscopic} fractions $\ket{\Psi^{mac}_i(t)}\equiv \big({\bf S}_i\ket{\vec k_0}\big)^{\otimes mN_t}$ (represented by the big solid slabs on the right)
become by  Eq.~(\ref{ortogonalizacja}) more and more distinguishable in the thermodynamic (\ref{thermod}) and the deep decoherence $t\gg\tau_D$ limits. 
Together with the decoherence mechanism (\ref{znikaogon}) this leads to 
a formation of the spectrum broadcast state (\ref{br2}) with pure environmental states, 
and hence to the objective existence of the (classical) state of the sphere.
\label{fig_ort}}
\end{center}
\end{figure} 

The asymptotic orthogonalization (\ref{nonoverlap}) is also straightforward to show in the case of pure environments.
The post-scattering states of the environment macro-fractions, Eq.~(\ref{rho_i}), are all pure:
\be
\varrho_i^{mac}(t)=\left({\bf S}_i\ket{\vec k_0}\bra{\vec k_0}{\bf S}_i^\dagger\right)^{\otimes mN_t}\equiv \ket{\Psi_i^{mac}(t)}\bra{\Psi_i^{mac}(t)},
\ee
so it is enough to consider their overlap:
\ben
&&\left|\langle\Psi_2^{mac}(t)|\Psi_1^{mac}(t)\rangle\right|=\left|\langle\vec k_0|{\bf S}_2^\dagger 
{\bf S}_1\vec k_0\rangle\right|^{L^2m\frac{N}{V}ct}\label{ortogonalizacja_0}\\
&&\xrightarrow{\text{therm.}}\text e^{-\frac{m}{\tau_D}t}.\label{ortogonalizacja}
\een
Thus, for $t\gg\tau_D$ the states of the macro-fractions $\Psi_i^{mac}(t)$ \emph{asymptotically orthogonalize} and moreover on the same timescale $\tau_D$ as
the decay of the coherent part described by Eq.~(\ref{ortogonalizacja}) (note that $0<m,f\leq 1$ so the timescales from Eqs.~(\ref{decay_ogon},\ref{ortogonalizacja})
do not differ considerably). This shows the asymptotic formation of the broadcast state (\ref{br2}) with pure encoding states $\varrho_i^{E_k}$:
\ben
&&\varrho_{S:fE}(0)=\varrho^S_0\otimes\underbrace{\varrho_0^{mac}\otimes\dots\otimes\varrho^{mac}_0}_{fM} 
\xrightarrow[\text{therm.}]{t\gg\tau_D}\varrho_{S:fE}(\infty)=\nonumber\\
&&\sum_{i=1,2}p_i\ket{\vec x_i}\bra{\vec x_i}
\otimes\underbrace{\ket{i^{mac}}\bra{i^{mac}}\otimes\dots\otimes\ket{i^{mac}}\bra{i^{mac}}}_{fM},\nonumber\\
\label{b-state}
\een
where  $p_i$ is given by Eq.~(\ref{pi}) and $\ket{i^{mac}}\equiv\ket{\Psi_i^{mac}(\infty)}$ emerges as the non-disturbing environmental basis
in the space of each macro-fraction, spanning a two-dimensional subspace, which  carries the correlation between the macro-fraction and the sphere
(this basis depends on the initial state $\ket{\vec k_0}$). Thus, the correlations become effectively among the qubits. 
The full process (\ref{b-state}) is a combination of the measurement of the system in the pointer basis
$\ket{\vec x_i}$ and spectrum broadcasting of the result, described by a CC-type channel \cite{my} :
\ben
\Lambda^{S\to fE}_\infty(\varrho_0^S)\equiv\sum_i\langle\vec x_i |\varrho^S_0\, \vec x_i\rangle\ket{i^{mac}}\bra{i^{mac}}^{\otimes fM}.
\label{L}
\een 
Entropic objectivity condition  and the classical plateau  
follow now form the Eq.~(\ref{b-state}):
\be\label{QD}
I[\varrho_{S:fE}(t)]\approx H_S,
\ee
because of the conditions (\ref{znikaogon},\ref{nonoverlap_norm})
(see the next Section for the details). Thus the mutual information 
becomes asymptotically independent of the fraction $f$ (as long as it is macroscopic).

In quantum Darwinism simulations for finite, fixed times $t$ (see e.g. Refs.~\cite{sfera_zurek,RiedelZurek}), one can observe that
the formation of the plateau is stronger driven by increasing the time rather than the macro-fraction $f$ (keeping all other parameters equal).
This can be straightforwardly explained by looking at the Eqs.~(\ref{decay_ogon},\ref{ortogonalizacja}): the fractions $f,m$ are by definition 
at most $1$, and hence have little effect on the decay of the exponential factors, while $t$ can be arbitrarily greater than $\tau_D$, thus 
accelerating the formation of the broadcast state (\ref{b-state}).

\subsection{Information-theoretical phases}\label{phases}
\begin{figure}[t]
\begin{center}
\includegraphics[scale=0.26]{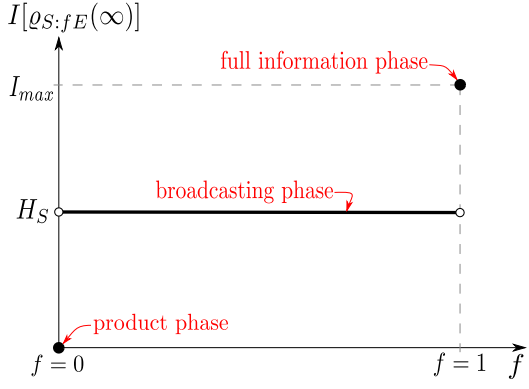}
\caption{(Color online) \emph{Information-theoretical phases of the sphere model (cf. \cite{sfera_zurek})}. 
Schematic phase diagram, showing three different phases of the illuminated sphere model, 
appearing in the thermodynamic and  the deep decoherence  $t\gg \tau_D$ limits. 
The horizontal axis is the observed fraction $f$ of the total photon number, understood modulo a microfraction. 
The vertical axis is the asymptotic mutual information between
the sphere $S$ and the fraction $fE$, $I[\varrho_{S:fE}(\infty)]$.
This is the limiting diagram for those from \cite{sfera_zurek} obtained for finite $t$. 
There are two phase transitions: At $f=0$ from the singular product phase (represented by the black point at zero) to
the broadcasting phase (the black line at $H_S$); and at $f=1$ from the broadcasting phase to the singular 
full information phase (black dot at $I_{max}$).
\label{phase}}
\end{center}
\end{figure}

There is a very distinct difference in  the macro- and microscopic behavior of the environment, already alluded to in Refs.~\cite{sfera_zurek,RiedelZurek} and summarized in Fig.~\ref{phase}.
>From Eq.(\ref{psipsi}) it follows that  within the sector (\ref{soft}) the post-scattering states of {\it individual} photons (\emph{micro-states}) $\ket{\Psi^{mic}_i}\equiv {\bf S}_i\ket{\vec k_0}$, 
become identical in the thermodynamic limit and hence encode no information about the sphere's localization:
\be\label{micro}
\langle\Psi^{mic}_2|\Psi^{mic}_1\rangle\equiv \langle\vec k_0|{\bf S}_2^\dagger {\bf S}_1\vec k_0\rangle\xrightarrow{\text{therm.}}1.
\ee
This is not surprising due to the condition (\ref{soft}). On the other hand, and despite of it, 
by  Eq.~(\ref{ortogonalizacja}) macroscopic groups of photons are able to resolve the sphere's position 
and in the asymptotic limit resolve it perfectly. 
It leads to the appearance of the different information-theoretical phases
in the model, which we now describe. We stress that the macro-fraction $m$
can be arbitrarily small (which only prolongs the orthogonalization time, cf. Eq.~(\ref{ortogonalizacja})), but
must scale with the total number of photons $N_t$. Indeed, for a microscopic, i.e. not scaling with $N_t$, fraction $\mu$
the limit (\ref{micro}) still holds: $[\langle\vec k_0|{\bf S}_2^\dagger {\bf S}_1\vec k_0\rangle]^\mu\xrightarrow{\text{therm.}}1$. 
Thus, if the observed portion of the environment is {\it microscopic}, the  
asymptotic post-scattering state is in fact a product one: 
\ben
&&\varrho_{S:\mu E}(0)=\varrho^S_0\otimes\left(\varrho_0^{mac}\right)^{\otimes \mu} \xrightarrow[\text{therm.}]{t\gg\tau_D}
\varrho_{S:\mu E}(\infty)=\nonumber\\
&&\sum_{i=1,2}p_i\ket{\vec x_i}\bra{\vec x_i}
\otimes\left({\bf S}_i\ket{\vec k_0}\bra{\vec k_0}{\bf S}_i^\dagger\right)^{\otimes \mu}=\\
&& \left(\sum_{i=1,2}p_i\ket{\vec x_i}\bra{\vec x_i}\right)
\otimes\ket{\Psi^{mic}}\bra{\Psi^{mic}}^{\otimes \mu},\label{product}
\een
where $\ket{\Psi^{mic}}\equiv {\bf S}_1\ket{\vec k_0} \cong {\bf S}_2\ket{\vec k_0}$ because of Eq.~(\ref{micro}) 
(and $\cong$ denotes  equality in the thermodynamic limit (\ref{thermod})). 
This is the product phase, in which $I[\varrho_{S:\mu E}(\infty)]=0$.

Conversely, if we have access to the full environment, ignoring perhaps only a microscopic fraction $\mu$,
the arguments leading to Eqs.~(\ref{decay_ogon},\ref{ortogonalizacja}) do not work anymore, since from Eq.~(\ref{micro}):
\be
\left|\tr {\bf S}_1\varrho^{ph}_0 {\bf S}_2^\dagger\right|^\mu \xrightarrow{\text{therm.}} 1,
\ee
and thus there is no decoherence nor orthogonalization. The post-scattering state 
contains then the full \emph{quantum} information about the system  due to the unsuppressed system-environment entanglement produced by the controlled-unitary
interaction (\ref{U}). As a result, the mutual information attains in the thermodynamical limit
its maximum value $I_{max}=2H_S$ (for a pure $\varrho_0^S$, since the interaction is
of a controlled unitary form (\ref{U})) and this defines the full information phase.
We note that the rise of $I_{S:fE}$ above $H_S$ certifies the presence of entanglement \cite{entropic}.
The intermediate phase described by Eq.~(\ref{b-state}) is the broadcasting phase (see Fig.~\ref{phase}).

The quantity experiencing discontinuous jumps is the mutual information between the system $S$ and the observed environment $fE$,
and the parameter which drives the phase transitions is the fraction size $f$.
As discussed above, each value of $f$ has to be understood modulo a micro-fraction. The appearance of the phase diagram
is a reflection of both the thermodynamic and the deep decoherence limits and its form is in agreement with the previously obtained
results (see e.g. Refs.~\cite{sfera_zurek,RiedelZurek}).

\subsection{Derivation of the entropic objectivity condition in the illuminated sphere model}\label{condition}
\label{IHS}
Here we present an independent derivation of the entropic objectivity condition
\be\label{QD2}
I[\varrho_{S:fE}(t)]\approx H_S,
\ee
for the illuminated sphere model.
Although illustrated on a concrete model, our derivation is indeed more general---instead of
a direct, asymptotic calculation of the mutual information $I[\varrho_{S:fE}(t)]$ in the model 
(cf. Refs.~\cite{ZwolakZurek,sfera_zurek,RiedelZurek}), we will show that Eq.~(\ref{QD2}) follows from the
mechanisms of: i) decoherence, Eq.~(\ref{znikaogon}), and ii) distinguishability, Eq.~(\ref{nonoverlap_norm}),  
once they are proven. In light of our findings, this puts a clear physical 
meaning to Eq.~(\ref{QD2})---it is a consequence of the state information broadcasting. 
Most of the proof is for general, mixed states.

Let the post-interaction $S:fE$ state for a fixed, finite box $L$ and time $t$ be $\varrho_{S:fE}(L,t)$.
It is given by Eqs.~(\ref{i=j},\ref{i ne j}) and now we 
explicitly indicate the dependence on $L$ in the notation. Then:
\ben
&&\left|H_S-I\left[\varrho_{S:fE}(L,t)\right]\right|\leq\nonumber\\
&&\left|I\left[\varrho_{S:fE}(L,t)\right]-I\left[\varrho^{i=j}_{S:fE}(L,t)\right]\right|\label{coh}\\
&&+\left|H_S-I\left[\varrho^{i=j}_{S:fE}(L,t)\right]\right|,\label{ort}
\een
where $\varrho^{i=j}_{S:fE}(L,t)$ is the decohered part of $\varrho_{S:fE}(L,t)$, given by Eq.~(\ref{i=j}).
We first bound the difference (\ref{coh}), decomposing the mutual information using conditional information 
$S_{\text{vN}}(\varrho_{S:fE}|\varrho_{fE})\equiv S_{\text{vN}}(\varrho_{S:fE})-S_{\text{vN}}(\varrho_{fE})$:
\be
I(\varrho_{S:fE})=S_{\text{vN}}\left(\varrho_S\right)-S_{\text{vN}}\left(\varrho_{S:fE}|\varrho_{fE}\right),
\ee
so that:
\ben
&&\left|I\left[\varrho_{S:fE}(L,t)\right]-I\left[\varrho^{i=j}_{S:fE}(L,t)\right]\right|\leq\nonumber\\
&&\left|S_{\text{vN}}\left[\varrho_{S}(L,t)\right]-S_{\text{vN}}\left[\varrho^{i=j}_{S}(L,t)\right]\right|+\label{SS}\\ 
&&\Big|S_{\text{vN}}\left[\varrho_{S:fE}(L,t)\big|\varrho_{fE}(L,t)\right]\nonumber\\
&&\quad\quad\quad\quad\quad -S_{\text{vN}}\left[\varrho_{S:fE}^{i=j}(L,t)\Big|\varrho_{fE}^{i=j}(L,t)\right]\Big|.
\label{Scond}
\een
>From Eq.~(\ref{soft}), the 
total $S:fE$ Hilbert space is finite-dimensional for a finite $L,t$: there are $fN_t=$$fL^2(N/V)ct$ photons (cf. Eq.~(\ref{Nt}))
and the number of modes of each photon is approximately $(4\pi/3)(L/2\pi\Delta x)^3$. 
Hence, the total dimension is $2\times L^2f(N/V)ct\times(1/6\pi^2)(L/\Delta x)^3<\infty$ and we can use the
Fannes-Audenaert \cite{FAd} and the Alicki-Fannes \cite{FannesAlicki} inequalities to bound
(\ref{SS}) and (\ref{Scond}) respectively (cf. Ref.~\cite{ZwolakZurek}). For (\ref{SS}) we obtain:
\ben
&&\left|S_{\text{vN}}\left[\varrho_{S}(L,t)\right]-S_{\text{vN}}\left[\varrho^{i=j}_{S}(L,t)\right]\right|\nonumber\\
&&\leq\frac{1}{2}\epsilon_E(L,t)\log(d_S-1)+h\left[\frac{\epsilon_E(L,t)}{2}\right],
\een
where $h(\epsilon)\equiv-\epsilon\log\epsilon-(1-\epsilon)\log(1-\epsilon)$ is the binary Shannon entropy
and:
\ben
&&\epsilon_E(L,t)\equiv ||\varrho_{S}(L,t)-\varrho^{i=j}_{S}(L,t)||_{tr}\\
&&=||\varrho^{i\ne j}_{S}(L,t)||_{tr}\cong 2|c_{12}|\left[1-\frac{1}{c\tau_DL^2}\left(\frac{N}{V}\right)^{-1}\right]^{L^2\frac{N}{V}ct}
\label{ELt}
\een
with $c_{12}\equiv\langle\vec x_1 |\varrho^S_0\, \vec x_2\rangle$, 
where we have used the reasoning (\ref{matrix}-\ref{decay_ogon}), 
but with $f=0$. For (\ref{Scond}) the same reasoning and the Alicki-Fannes inequality give:
\ben
&&\left|S_{\text{vN}}\left[\varrho_{S:fE}(L,t)\big|\varrho_{fE}(L,t)\right]
-S_{\text{vN}}\left[\varrho_{S:fE}^{i=j}(L,t)\big|\varrho_{fE}^{i=j}(L,t)\right]\right|\nonumber\\
&&\leq4\epsilon_{fE}(L,t)\log d_S+2 h\left[\epsilon_{fE}(L,t)\right],
\een
with:
\ben
\epsilon_{fE}(L,t)&\equiv& ||\varrho_{S:fE}(L,t)-\varrho^{i=j}_{S:fE}(L,t)||_{tr}\\
&=&||\varrho^{i\ne j}_{S:fE}(L,t)||_{tr}\\
&\cong& 2|c_{12}|\left[1-\frac{1}{c\tau_DL^2}\left(\frac{N}{V}\right)^{-1}\right]^{L^2(1-f)\frac{N}{V}ct}.\label{EfLt}
\een
Above $L,t$ are big enough so that $\epsilon_E(L,t), \epsilon_{fE}(L,t)<1$. 
Eqs.~(\ref{SS}-\ref{EfLt}) give an upper bound on the difference (\ref{coh}) in terms of the 
decoherence speed (\ref{znikaogon}).

To bound the "orthogonalization" part (\ref{ort}) (see Ref.~\cite{ZwolakZurek} for a related analysis), 
we note that since $\varrho^{i=j}_{S:fE}(L,t)$ is a CQ-state
(cf. Eq.~(\ref{i=j})), its mutual information is given by the Holevo quantity \cite{Holevo}:
\ben
I\left[\varrho^{i=j}_{S:fE}(L,t)\right]=\chi\left\{p_i,\varrho_i^{mac}(t)^{\otimes fM}\right\},
\een
where $p_i$ is given by Eq.~(\ref{pi}).
>From the Holevo Theorem it is bounded by \cite{Holevo}:
\be\label{hol}
I_{max}(t)\leq\chi\left\{p_i,\varrho_i^{mac}(t)^{\otimes fM}\right\}\leq H\left(\{p_i\}\right)\equiv H_S,
\ee
where $I_{max}(t)\equiv \max_{\mathcal E}I[p_i\pi^{\mathcal E}_{j|i}(t)]$ is the fixed time maximal mutual information, extractable 
through generalized measurements $\{\mathcal E_j\}$ on the  ensemble 
$\{p_i,\varrho_i^{mac}(t)^{\otimes fM}\}$, and the conditional probabilities read:
\be\label{piE}
\pi^{\mathcal E}_{j|i}(t)\equiv\tr[\mathcal E_j\varrho_i^{mac}(t)^{\otimes fM}]
\ee
(here and below $i$ labels the states, while $j$ the measurement outcomes).
We now relate $I_{max}(t)$ to the generalized overlap   
$B\left[\varrho_1^{mac}(t)^{\otimes fM},\varrho_2^{mac}(t)^{\otimes fM}\right]$ (cf. Eq.~(\ref{nonoverlap_norm})),
which we have calculated for pure states in Eq.~(\ref{ortogonalizacja_0},\ref{ortogonalizacja}). Using the method of Ref.~\cite{Fuchs},
slightly modified to unequal a priori probabilities $p_i$, we obtain for an arbitrary measurement $\mathcal E$:
\ben
&&I\left(\pi^{\mathcal E}_{j|i}p_i\right)=I\left(\pi^{\mathcal E}_{i|j}\pi^{\mathcal E}_j\right)
=H\left(\{p_i\}\right)-\sum_{j=1,2}\pi^{\mathcal E}_jh\left(\pi^{\mathcal E}_{1|j}\right)\nonumber\\
&&\\
&&\geq H\left(\{p_i\}\right)-2\sum_{j=1,2}\pi^{\mathcal E}_j \sqrt{\pi^{\mathcal E}_{1|j}\left(1-\pi^{\mathcal E}_{1|j}\right)}\\
&&=H\left(\{p_i\}\right)-2\sqrt{p_1p_2}\sum_{j=1,2}\sqrt{\pi^{\mathcal E}_{j|1}\pi^{\mathcal E}_{j|2}},
\een
where we have first used Bayes Theorem $\pi^{\mathcal E}_{i|j}=(p_i/\pi^{\mathcal E}_j)\pi^{\mathcal E}_{j|i}$,
$\pi^{\mathcal E}_j\equiv\sum_i\pi^{\mathcal E}_{j|i}p_i=\tr(\mathcal E_j\sum_i\varrho_i)$, then the fact that we have only two states:
$\pi^{\mathcal E}_{2|j}=1-\pi^{\mathcal E}_{1|j}$, so that $H(\pi^{\mathcal E}_{\cdot|j})=h(\pi^{\mathcal E}_{1|j})$,
and finally $h(p)\leq2\sqrt{p(1-p)}$. On the other hand, 
$B(\varrho_1,\varrho_2)=\min_{\mathcal E}\sum_j\sqrt{\pi^{\mathcal E}_{j|1}\pi^{\mathcal E}_{j|2}}$ \cite{Fuchs}. Denoting the optimal
measurement by $\mathcal E_*^B(t)$ and recognizing that $H(\{p_i\})=H_S$, we obtain:
\ben
&&I_{max}(t)\geq I\left[p_i\pi^{\mathcal E_*^B(t)}_{j|i}(t)\right]\geq H_S-\\
&&-2\sqrt{p_1p_2}\,B\left[\varrho_1^{mac}(t)^{\otimes fM},\varrho_2^{mac}(t)^{\otimes fM}\right]\\
&&=H_S-2\sqrt{p_1p_2}\,B\left[\varrho^{mac}_1(t),\varrho^{mac}_2(t)\right]^{fM} 
\een
Inserting the above into the bounds (\ref{hol}) gives the desired upper bound on the difference (\ref{ort}):
\ben
&&\left|H_S-I\left[\varrho^{i=j}_{S:fE}(L,t)\right]\right|\leq
2\sqrt{p_1p_2}\,B\left[\varrho^{mac}_1(t),\varrho^{mac}_2(t)\right]^{fM}\nonumber\\
&&
\een
where the generalized overlap is given by Eqs.~(\ref{ortogonalizacja_0},\ref{ortogonalizacja}): 
\ben
&&B\left[\varrho^{mac}_1(t),\varrho^{mac}_2(t)\right]=\left|\langle\Psi_2^{mac}(t)|\Psi_1^{mac}(t)\rangle\right|\cong\nonumber\\
&&\quad\quad\quad\left[1-\frac{1}{c\tau_D L^2}\left(\frac{N}{V}\right)^{-1}\right]^{L^2m\frac{N}{V}ct}.
\label{Bh}
\een

Gathering all the above facts together finally leads to a bound on   
$\left|H_S-I\left[\varrho_{S:fE}(L,t)\right]\right|$ in terms of the speed of 
i) decoherence (\ref{znikaogon}) and ii) distinguishability (\ref{nonoverlap_norm}):
\ben
&&\left|H_S-I\left[\varrho_{S:fE}(L,t)\right]\right|\leq h\left[\frac{\epsilon_E(L,t)}{2}\right]+
 2h\left[\epsilon_{fE}(L,t)\right]+\nonumber\\
&&\label{gen1}\\
&&4\epsilon_{fE}(L,t)\log 2+2\sqrt{p_1p_2}\,B\left[\varrho^{mac}_1(t),\varrho^{mac}_2(t)\right]^{fM},\label{gen2}
\een
where $\epsilon_E(L,t)$, $\epsilon_{fE}(L,t)$, $B\left[\varrho^{mac}_1(t),\varrho^{mac}_2(t)\right]$
are given by Eqs.~(\ref{ELt}), (\ref{EfLt}), and (\ref{Bh}) respectively. Choosing
$L,t$ big enough so that $\epsilon_E(L,t),\epsilon_{fE}(L,t)\leq 1/2$ (when the binary 
entropy $h(\cdot)$ is monotonically increasing), we remove the unphysical box and obtain an estimate on
the speed of convergence of $I\left[\varrho_{S:fE}(L,t)\right]$ to $H_S$:
\ben
&&\lim_{L\to\infty}\left|H_S-I\left[\varrho_{S:fE}(L,t)\right]\right|\leq 
h\left(|c_{12}|\text e^{-\frac{t}{\tau_D}}\right)\\
&&+2h\left(2|c_{12}|\text e^{-\frac{(1-f)}{\tau_D}t}\right)
+8|c_{12}|\text e^{-\frac{(1-f)}{\tau_D}t}\log 2\\
&&+2\sqrt{p_1p_2}\text e^{-\frac{f}{\tau_D}t}.
\een
This finishes the derivation of the condition (\ref{QD2}).

We note that the result (\ref{gen1},\ref{gen2}) is in fact a general statement, valid 
in any model where: i) the system $S$ is effectively a qubit; ii) the system-environment
interaction is of a environment-symmetric controlled-unitary type:
\begin{lemma}
Let a two-dimensional quantum system $S$ interact with $N$ identical environments, each described by
a $d$-dimensional Hilbert space, through a controlled-unitary interaction:
\be
U(t)\equiv\sum_{i=1,2}\ket i\bra i\otimes U_i(t)^{\otimes N}.
\ee
Let the initial state be $\varrho_{S:E}(0)=\varrho_0^S\otimes(\varrho_0^E)^{\otimes N}$ 
and $\varrho_{S:E}(t)\equiv U(t)\varrho_{S:E}(0)U(t)^\dagger$. Then for any $0<f<1$ and $t$ big enough:
\ben
&&\left|H(\{p_i\})-I\left[\varrho_{S:fE}(t)\right]\right|\leq h\left[\frac{\epsilon_E(t)}{2}\right]+
 2h\left[\epsilon_{fE}(t)\right]+\nonumber\\
&&\\
&&4\epsilon_{fE}(t)\log 2+2\sqrt{p_1p_2}\,B\left[\varrho_1(t),\varrho_2(t)\right]^{fN},\label{gen3}
\een
where: 
\ben
&&p_i\equiv\langle i|\varrho_0^S|i\rangle,\,\varrho_i(t)\equiv U_i(t)\varrho_0^EU_i(t)^\dagger,\\
&&\epsilon_E(t)\equiv ||\varrho_{S}(t)-\varrho^{i=j}_{S}||_{tr},\\
&&\epsilon_{fE}(t)\equiv ||\varrho_{S:fE}(t)-\varrho^{i=j}_{S:fE}(t)||_{tr}.
\een
\end{lemma}

\end{document}